\documentstyle[prl,aps,twocolumn,eqsecnum,epsf]{revtex}
\begin{document}
\draft
\title{Path-integral Monte Carlo Simulations without the Sign Problem:
Multilevel Blocking Approach for Effective Actions}
\author{R.~Egger$^1$, L.~M{\"u}hlbacher$^{1,2}$, and C.H.~Mak$^2$}
\address{
${}^1$~Fakult\"at f\"ur Physik, Albert-Ludwigs-Universit\"at,  
 D-79104 Freiburg, Germany\\
${}^2$~Department of Chemistry, University of Southern California,
Los Angeles, CA 90089-0482}
\date{Date: \today}
\maketitle
\begin{abstract}
The multilevel blocking algorithm recently proposed as a
possible solution to the sign problem in 
path-integral Monte Carlo simulations has been extended to systems 
with long-ranged interactions along the Trotter direction.
As an application, new results for the real-time quantum dynamics of the 
spin-boson model are presented.
\end{abstract}

\narrowtext

\section{Introduction}

Path-integral Monte Carlo (PIMC) simulations are useful for 
extracting exact results on many-body quantum systems \cite{qmc}.   
In principle, PIMC methods can be used to study both equilibrium 
as well as dynamical problems.  But in the cases of 
fermions and real-time dynamics, PIMC suffers 
from the notorious ``sign problem'' which renders such simulations  
unstable.  This sign problem manifests itself as 
an exponential decay of the 
signal-to-noise ratio for large systems or long real times 
\cite{ceperley,sign,berne}.  Its origin is at the heart of quantum
mechanics itself, namely the interference of different quantum paths
contributing to the path integral might be destructive 
due to exchange effects or due to the oscillatory nature
of the real-time evolution operator.  
Besides approximate treatments \cite{ceperley} the sign problem has
remained unsolved.  

Very recently, a new strategy has 
been proposed as a possible approach to a complete solution of 
the sign problem.   This so-called multi-level blocking (MLB)
algorithm \cite{mlb,mlbrt} is 
a systematic implementation of the simple blocking
idea --- by sampling ``blocks'' instead of single paths,
one can {\em always} reduce the sign problem \cite{egger94}.
Defining a suitable hierarchy of blocks by
grouping them into different ``levels'', crucial information
about the phase cancellations among different quantum paths 
can then be recursively transferred from the bottom to the top level.
Given sufficient 
computer memory, such an approach was shown to be able to eliminate
the sign problem in a stable and exact manner \cite{mlb}.
But to date, the MLB algorithm has only been formulated to 
solve the sign problem in PIMC simulations with 
nearest-neighbor interactions along the Trotter direction.
This situation is encountered under a direct Trotter-Suzuki breakup of the
short-time propagator.

In this paper, we report an extension of the MLB approach
to the case of effective actions that may include arbitrarily long-ranged
interactions.   Such effective actions that are non-local in Trotter time
may arise from degrees of freedoms having been traced out, 
e.g., a harmonic heat bath \cite{leggett},
or through a Hubbard-Stratonovich transformation, e.g., in 
auxiliary-field MC simulations of lattice fermions \cite{sign}.
Remarkably, because such effective
actions capture much of the physics, e.g., symmetries or
the dissipative influence of the traced-out degrees of freedom, 
the corresponding path integral very often exhibits a
significantly reduced ``intrinsic'' sign problem compared to the original
(time-local) formulation. 
The present generalization of the MLB algorithm was developed
to take advantage of this fact.  We note that in a PIMC simulation
with only nearest-neighbor interactions along the Trotter direction,
the original MLB approach  \cite{mlb} is more efficient than the method described
below, which therefore should be used only for time-non-local actions. 

To be specific, we focus on the dynamical sign problem
arising in real-time PIMC computations here.  
The modifications required to implement the method for
fermion simulations are then straightforward. 
The structure of this paper is as follows.  In Sec.~\ref{gen}
the general strategy to deal with long-ranged interactions
in a MLB scheme is outlined.  A detailed exposition of the computational
method can be found in Sec.~\ref{mlbsim}.  
We have studied the real-time dynamics of the celebrated spin-boson 
system \cite{leggett} using this approach.  Details about this
application, performance issues related to the sign problem,
and numerical results   are presented in Sec.~\ref{spinboson}.
Finally, Sec.~\ref{conc} offers some conclusions.  

\section{General considerations} \label{gen}

We consider a discretized path integral along a certain contour in 
the complex-time plane.  In a typical real-time calculation, 
there is a forward branch from $t=0$ to $t=t^*$, 
where $t^*$ is the maximum time studied in the simulation, 
followed by a branch going back to the origin, 
and then by an imaginary-time branch 
from $t=0$ to $t=-i\hbar \beta$. We focus on
a ``factorized'' initial preparation where the relevant degrees of freedom,
denoted by $\bbox{r}(t)$,  are held fixed for $t<0$ \cite{leggett,foot1}.  
That implies that the imaginary-time dynamics must be
frozen at the corresponding value, and we only need to 
sample on the two real-time branches.  Note that such
a nonequilibrium calculation cannot proceed in a 
standard way by first doing an imaginary-time QMC simulation 
followed by analytic continuation of the numerical data \cite{qmc}.
The quantum numbers $\bbox{r}(t)$ at a 
given time may be discrete or continuous variables.

Using time slices of length $t^*/P$,  we combine
forward [$\bbox{r}(t_m)$] and backward [$\bbox{r}'(t_m)$]
path configurations at time $t_m=m t^*/P$
into the configuration $\bbox{s}_m$, where $m=1,\ldots,P$.
The configuration at $t=0$ is held fixed,
and for $t=t^*$ we must be in a diagonal state, $\bbox{r}(t^*)
=\bbox{r}'(t^*)$.  For an efficient application of the current 
method, it is essential to combine several neighboring
slices $m$ into new ``blocks''.  For instance, think of $m=1,\ldots,5$
as a new ``slice'' $\ell=1$, $m=6,\ldots,10$ as another slice $\ell=2$,
and so on.  Combining $q$ elementary slices
into a block $\bbox{s}_\ell$, instead of the original $P$ slices we have
$L=P/q$ blocks, where $L$ is the number of MLB ``levels''.
In actual applications, there is considerable freedom in 
how these blocks are defined, e.g. if 
there is hardly any intrinsic sign problem, or if there
are only few variables in $\bbox{r}$, one may choose
larger values of $q$. 
Additional flexibility can be gained
by choosing different $q$ for different blocks.

Say we are interested in sampling the configurations $\bbox{s}_L$ on the top
level $\ell=L$ according to the appropriate
matrix elements of the (reduced) density matrix,
\begin{equation} \label{rho}
\rho(\bbox{s}_L) = Z^{-1}\sum_{ \bbox{s}_1,\ldots,
\bbox{s}_{L-1}} \exp\{- S[\bbox{s}_1, \ldots,\bbox{s}_L] \} \;,
\end{equation}
where $S$ is the effective action under study and
$Z$ is a normalization constant so that 
\begin{equation}\label{norm}
\sum_{\bbox{s}_L} \rho(\bbox{s}_L) = 1\;.
\end{equation}
Due to the time-non-locality of this action, there will be interactions
among all blocks $\bbox{s}_\ell$.  The
sum in Eq.~(\ref{rho}) denotes either an integration over continuous
degrees of freedom or a discrete sum. In the case
of interest here, the effective action is complex-valued and
$e^{-S}/|e^{-S}|$
represents an oscillatory phase factor ($\pm 1$ for the fermion sign problem).
The ``naive approach'' to the
sign problem is to sample configurations using the 
positive definite weight function
\begin{equation} \label{standard}
{\cal P} \sim |\exp\{-S\}| \;,
\end{equation}
and to include the oscillatory phase in the accumulation
procedure.  Precisely this leads to the exponentially
fast decay of the signal-to-noise ratio with $t^*$.

The proposed MLB simulation 
scheme starts by sampling on the finest level $\ell=1$,
so only variables in the first block corresponding to $m=1,\ldots,q$
are updated.  During this procedure, interference among different
paths will take place.  Since only relatively few degrees of freedom
are sampled, however, the resulting interference information can be
quantified in a  controlled way by employing
so-called ``level-$\ell$ bonds'' (here $\ell=1$).
As long as $q$ is chosen sufficiently small, the interference cannot
lead to numerical instabilities, and the 
sign cancellations occuring while sampling on level
$\ell=1$ can thus be synthesized and
transferred to the level $\ell=2$, where the sampling is carried out next. 
Here the procedure is repeated, and by proceeding recursively 
up to the top level $\ell=L$,
this strategy can eliminate the sign problem.   The main bottleneck of the
method comes from the immense memory requirements, since one needs
to store and update the level-$\ell$ bonds on all levels during the 
Monte Carlo sampling (see below for details).  To summarize, the main
idea of our approach is to subdivide the allowed interferences among the 
quantum paths into small subunits (blocks) such that no sign problem occurs
when (stochastically) summing over the paths within each subunit.  
The basic observation underlying our method is therefore almost trivial: The
sign problem does not occur in a sufficiently small system.
The nontrivial computational task
then consists of bringing together the interference signals from different
blocks, which is done by recursively forming blocks on subsequent higher levels.

Instead of the ``circular'' structure of the time contour inherent in the trace
operation, it is actually more helpful
to view the problem as a linear chain, where the proposed
MLB scheme proceeds from left to right.  In the case of local actions
with only nearest-neighbor interactions along Trotter time, a different
recursion scheme was implemented in Refs.\cite{mlb,mlbrt} which is close
in spirit to the usual block-spin transformations used in renormalization
group treatments of spin chains.  For both MLB implementations,
however, the underlying blocking idea is identical, 
and the non-locality of the effective action studied here
only requires one to abandon block-spin-like
transformations in favor of the ``moving-along-the-chain'' picture. 

Below we assume that one can decompose the effective action
according to
\begin{equation} \label{decompos}
S[\bbox{s}_1, \ldots,\bbox{s}_L]  = \sum_{\ell=1}^L
W_\ell[\bbox{s}_\ell,\ldots,\bbox{s}_L] \;.
\end{equation}
All dependence on a configuration $\bbox{s}_\ell$ is then
contained in the ``partial actions'' $W_{\lambda}$ with
$\lambda\leq \ell$.   
One could, of course, put all $W_{\ell>1}=0$,
but the approach becomes more powerful if 
a nontrivial decomposition is possible.  

\section{Multilevel blocking approach}\label{mlbsim}

In the following, we describe in detail how the  MLB algorithm 
for effective actions is implemented in practice.  
The MC sampling starts on the finest level $\ell=1$, where
only the configuration $\bbox{s}_{\ell=1}$ containing the elementary slices
$m=1,\ldots,q$ will be updated with all  $\bbox{s}_{\ell>1}$ remaining fixed
at their initial values $\bbox{s}_\ell^0$.
Using the weight function
\[
{\cal P}_0[\bbox{s}_1] =  |\exp\{-W_1[\bbox{s}_1,
\bbox{s}_2^0, \ldots, \bbox{s}_L^0]\}| \;,
\]
we generate $K$ samples $\bbox{s}_1^{(i)}$,
where $i=1,\ldots,K$,  and store them for later use. 
To effectively solve the sign problem and 
to avoid a bias in the algorithm, 
the sample number $K$ should be chosen large enough, see below
and Ref.\cite{mlb}. 
For $K=1$, the algorithm simply reproduces the naive approach.

The stored samples are now employed to generate information about
the sign cancellations.  All knowledge about the interference that occured
at this level is encapsulated in the quantity
\begin{eqnarray} \label{l11} 
B_1 &=& \left\langle\frac{ 
  \exp\{- W_1[\bbox{s}_1, \ldots, \bbox{s}_L]\}
                           }{
  |\exp\{- W_1[\bbox{s}_1, \bbox{s}_2^0, \ldots, \bbox{s}_L^0]\}|
                            } \right\rangle_{{\cal P}_0[\bbox{s}_1]}\\
\nonumber
 &=& C_0^{-1} \sum_{\bbox{s}_1} 
\exp\{- W_1[\bbox{s}_1, \ldots, \bbox{s}_L]\}\\
\nonumber
 &=& K^{-1} \sum_{i=1}^K \frac{ 
\exp\{- W_1[\bbox{s}_1^{(i)}, \bbox{s}_2, \ldots, \bbox{s}_L]\}
                                     }{
|\exp\{- W_1[\bbox{s}_1^{(i)}, \bbox{s}_2^0, \ldots, \bbox{s}_L^0]\}|
                                      } \\ \nonumber
&=& B_1[\bbox{s}_2,\ldots,\bbox{s}_L] \;,
\end{eqnarray}
which we call ``level-1 bond'' in analogy to Ref.\cite{mlb},
with the normalization constant
$C_0=\sum_{\bbox{s}_1} {\cal P}_0[\bbox{s}_1]$.
The third line follows by noting
that the $\bbox{s}_1^{(i)}$ were generated according
to the weight ${\cal P}_0$. 
This equality requires that $K$ is sufficiently large and
that $q$ is sufficiently small in order to 
provide a good statistical estimate of the level-1 bond.

Combining the second expression in Eq.~(\ref{l11}) with
Eq.~(\ref{rho}), we rewrite the
density matrix in the following way:
\begin{eqnarray}  \label{rhos}
\rho(\bbox{s}_L) &=& Z^{-1}\sum_{ \bbox{s}_2,\ldots , \bbox{s}_{L-1}}
\exp\left \{-\sum_{\ell>1}W_\ell \right \} 
C_0 B_1 \\
 &=& Z^{-1}\sum_{ \bbox{s}_1,\ldots , \bbox{s}_{L-1}} 
{\cal P}_0 B_1 \prod_{\ell>1} e^{-W_\ell} \;. \nonumber
\end{eqnarray}
When comparing Eq.~(\ref{rhos}) with Eq.~(\ref{rho}), we
see that the entire sign problem has now formally been transferred to
levels $\ell>1$, since oscillatory phase factors only 
arise when sampling on these higher levels.
  Note that  $B_1=B_1[\bbox{s}_2,\ldots,\bbox{s}_L]$
introduces couplings among {\em all} levels $\ell>1$, 
in addition to the ones 
already contained in the effective action $S$.

We now proceed to the next level $\ell=2$ and, according to
Eq.~(\ref{rhos}), update
configurations for $m=q+1,\ldots,2q$ using the weight
\begin{equation}
{\cal P}_1[\bbox{s}_2] =
|B_1[\bbox{s}_2, \bbox{s}_3^0, \ldots, \bbox{s}_L^0]
\exp\{-W_2[\bbox{s}_2, \bbox{s}_3^0, \ldots, \bbox{s}_L^0]\}| \;.
\end{equation}
Under the move $\bbox{s}_2\to \bbox{s}_2'$, we
should then resample and update the level-1 bonds, $B_1\to B_1'$.
Exploiting the fact
that the stored $K$ samples $\bbox{s}_1^{(i)}$ are correctly 
distributed for the original configuration $\bbox{s}_2^0$, 
the updated bond can be computed  according to
\begin{equation}\label{bupdate}
B_1^\prime= K^{-1}\sum_{i=1}^K \frac{ 
\exp\{- W_1[\bbox{s}_1^{(i)}, \bbox{s}_2', \ldots, \bbox{s}_L]\}
                                    }{
|\exp\{- W_1[\bbox{s}_1^{(i)}, \bbox{s}_2^0, \ldots, \bbox{s}_L^0]\}|
                                     } \;.
\end{equation}
Again, to obtain
an accurate estimate for $B_1^\prime$, the number $K$ should be
sufficiently large.
In the end, sampling under the weight ${\cal P}_1$
implies that the probability for accepting the move $\bbox{s}_2\to 
\bbox{s}_2^\prime$
under the Metropolis algorithm is
\begin{equation}
p = \left| \frac{\sum_i  \frac{\exp\{-W_1[\bbox{s}_1^{(i)},\bbox{s}'_2,
\bbox{s}_3^0,
\ldots]\}}{|\exp\{-W_1[\bbox{s}_1^{(i)},\bbox{s}_2^0,\ldots]\}|}}
{\sum_i  \frac{\exp\{-W_1[\bbox{s}_1^{(i)},\bbox{s}_2,\bbox{s}_3^0,
\ldots]\}}{|\exp\{-W_1[\bbox{s}_1^{(i)},\bbox{s}_2^0,\ldots]\}|}} \right |
\times \left| \frac{ \exp\{-W_2[\bbox{s}'_2,\bbox{s}_3^0,\ldots]\}}
{\exp\{-W_2[\bbox{s}_2,\bbox{s}_3^0,\ldots]\}} \right | \;.
\end{equation}

Using this method, we generate $K$ samples 
$\bbox{s}_2^{(i)}$, store them,
and compute the level-2 bonds, 
\begin{eqnarray} \label{bond2}
B_2 &=& \left\langle \frac{
B_1[\bbox{s}_2, \bbox{s}_3,\ldots]
\exp\{- W_2[\bbox{s}_2, \bbox{s}_3,\ldots]\}
                          }{
|B_1[\bbox{s}_2, \bbox{s}_3^0, \ldots]
\exp\{- W_2[\bbox{s}_2, \bbox{s}_3^0, \ldots]\} |
                           } \right\rangle_{{\cal P}_1[\bbox{s}_2]}\\ \nonumber
 &=& C_1^{-1} \sum_{\bbox{s}_2} B_1[\bbox{s}_2,\ldots]
 \exp\{- W_2[\bbox{s}_2, \ldots]\} \\ \nonumber
&=& K^{-1} \sum_{i=1}^K \frac{ 
B_1[\bbox{s}_2^{(i)}, \bbox{s}_3, \ldots]
\exp\{- W_2[\bbox{s}_2^{(i)}, \bbox{s}_3, \ldots]\}
                                    }{
|B_1[\bbox{s}_2^{(i)},\bbox{s}_3^0, \ldots]
\exp\{- W_2[\bbox{s}_2^{(i)}, \bbox{s}_3^0, \ldots]\}|
                                     } \\ \nonumber
&=& B_2[\bbox{s}_3, \ldots, \bbox{s}_L] \;,
\end{eqnarray}
with $C_1 = \sum_{\bbox{s}_2} {\cal P}_1[\bbox{s}_2]$. 
Following our above strategy, we then rewrite the
reduced density matrix by combining Eq.~(\ref{rhos}) and  the 
second line of Eq.~(\ref{bond2}).
This yields
\begin{eqnarray}\label{rhos2}
\rho(\bbox{s}_L) &=& 
Z^{-1} \sum_{\bbox{s}_3, \ldots, \bbox{s}_{L-1}}
\exp\left\{ - \sum_{\ell>2} W_\ell \right\}
C_0 C_1 B_2 \\ \nonumber
&=& Z^{-1} \sum_{\bbox{s}_1, \ldots, \bbox{s}_{L-1}} 
{\cal P}_0 {\cal P}_1 B_2 \prod_{\ell>2} e^{-W_\ell} \;.
\end{eqnarray}
Clearly, the sign problem has been transferred one block further
to the right along the chain. Note that the normalization
constants $C_0, C_1,\ldots$ depend only on the initial configuration
$\bbox{s}_\ell^0$ so that their precise values need not be known.

This procedure is now iterated in a recursive manner.
Sampling on level $\ell$ using the 
weight function
\begin{equation}
{\cal P}_{\ell-1}[\bbox{s}_\ell] =
|B_{\ell-1}[\bbox{s}_\ell, \bbox{s}_{\ell+1}^0, \ldots]
 \exp\{-W_\ell[\bbox{s}_\ell, \bbox{s}_{\ell+1}^0, \ldots]\}|
\end{equation}
requires the recursive update of all bonds $B_{\lambda}$
with $\lambda<\ell$. Starting with $B_1\to B_1'$ and putting $B_0=1$,
this recursive update is done according to
\begin{eqnarray} \label{brec}
\lefteqn{B'_{\lambda} = K^{-1} } \\
 && \times\sum_{i=1}^K \frac{ 
B'_{\lambda-1}[\bbox{s}_\lambda^{(i)}, \bbox{s}_{\lambda+1}, \ldots] \exp\{-
W'_\lambda[\bbox{s}_\lambda^{(i)}, \bbox{s}_{\lambda+1}, \ldots]\}
                                    }{
|B_{\lambda-1}[\bbox{s}_\lambda^{(i)},
\bbox{s}_{\lambda+1}^0, \ldots]\exp\{- 
W_\lambda[\bbox{s}_\lambda^{(i)}, \bbox{s}_{\lambda+1}^0, \ldots]\}|
                                     } \;, \nonumber
\end{eqnarray}
where the primed bonds or partial actions depend on $\bbox{s}'_{\ell}$
and the unprimed ones on $\bbox{s}^0_\ell$.
Iterating this to get the updated bonds $B_{\ell-2}$ for all 
$\bbox{s}_{\ell-1}^{(i)}$,   
the test move $\bbox{s}_\ell\to \bbox{s}'_\ell$
is then accepted or rejected according to the probability
\begin{equation} \label{prob}
p = \left| \frac{B_{\ell-1}[\bbox{s}'_\ell, \bbox{s}_{\ell+1}^0, \ldots]
\exp\{-W_\ell[\bbox{s}'_\ell, \bbox{s}_{\ell+1}^0, \ldots]\}}
{B_{\ell-1}[\bbox{s}_\ell, \bbox{s}_{\ell+1}^0, \ldots]
\exp\{-W_\ell[\bbox{s}_\ell, \bbox{s}_{\ell+1}^0, \ldots]\}} \right | \;.
\end{equation}
On this level, 
we again generate $K$ samples $\bbox{s}_{\ell}^{(i)}$, store them
and compute the level-$\ell$ bonds according to
\begin{eqnarray}
\lefteqn{B_{\ell}[\bbox{s}_{\ell+1},\ldots] = K^{-1}} \\
 && \times \sum_{i=1}^K \frac{ 
B_{\ell-1}[\bbox{s}_\ell^{(i)}, \bbox{s}_{\ell+1}, \ldots]
\exp\{- W_\ell[\bbox{s}_\ell^{(i)}, \bbox{s}_{\ell+1}, \ldots]\}
                      }{
|B_{\ell-1}[\bbox{s}_\ell^{(i)}, \bbox{s}_{\ell+1}^0, \ldots]
\exp\{- W_\ell[\bbox{s}_\ell^{(i)}, \bbox{s}_{\ell+1}^0, \ldots]\}|
                       }\;. \nonumber
\end{eqnarray}
This process is iterated up to the top level,
where the observables of interest may be computed.

Since the sampling of $B_{\ell}$ requires the resampling of 
all lower-level bonds, the memory and CPU requirements of the
algorithm laid out here are quite large.  
For $\lambda<\ell-1$, one needs to update $B_{\lambda}\to B'_{\lambda}$ for
all $\bbox{s}_{\ell'}^{(i)}$ with $\lambda< \ell' < \ell$, 
which implies a tremendous amount of computer memory and CPU time,
scaling approximately $\sim K^L$ at the top level. 
Fortunately, an enormous simplification can often be achieved 
by exploiting the fact that the interactions
among distant slices are usually weaker than between near-by
slices.  For instance, when updating level $\ell=3$,
the correlations with the configurations $\bbox{s}_1^{(i)}$
may be very weak, and instead of summing over all $K$ samples
$\bbox{s}_1^{(i)}$ in the update of the bonds $B_{\lambda<\ell}$,
we may select only a small subset.  
When invoking this argument, one should be careful to 
also check that the additional interactions coming from
the level-$\lambda$ bonds with $\lambda<\ell$ are sufficiently
short-ranged.  From the definition of these bonds, this
is to be expected though.

Remarkably, this algorithm can significantly relieve the severity
of the sign problem.
Let us first give a simple qualitative argument supporting this statement
for the original MLB method of Ref.\cite{mlb}, where $P=2^L$ with
$L$ denoting the number of levels.
If one needs $K$ samples for each slice on a given level in order
to have satisfactory statistics despite of the sign problem,
the total number of paths needed in the naive approach
depends exponentially on $P$, namely  $\sim K^P$. 
This is precisely the well-known exponential
severity of the sign problem under the 
naive approach.  However, with MLB the work on the last
level [which is the only one affected by a sign problem provided $K$
was chosen sufficiently large] is only $\sim K^L$.
 So in MLB, the work needed to sample the $K^P$
paths with satisfactory statistical accuracy
grows $\sim K^{\log_2 P} = P^{\log_2 K}$, i.e.,
only algebraically with $P$.
Provided the interactions along the Trotter time decay 
sufficiently fast, a similar qualitative 
argument can be given for  the generalized
MLB algorithm proposed here. For the application 
described below, we have  
indeed found only algebraic dependences of the required CPU times
and memory resources with the maximum real time $t^*$, 
instead of exponential ones as encountered in the naive approach.
Further details of the simulation procedure are provided in
the next section.

\section{Application: Spin-boson dynamics}
\label{spinboson}

To demonstrate this MLB algorithm for path integral simulations with
long-range interactions in the Trotter direction, we 
study the real-time dynamics of the spin-boson model,
\begin{eqnarray} \label{spbos}
H &=& - (\hbar\Delta/2)\, \sigma_x + (\hbar\epsilon/2)\, \sigma_z  \\
\nonumber &+& \sum_\alpha
\left[ \frac{p_\alpha^2}{2m_\alpha} +
{\textstyle \frac{1}{2}}  m_\alpha \omega_\alpha^2
\left(x_\alpha - \frac{c_\alpha}{m_\alpha \omega_\alpha^2} \sigma_z\right)^2
\right] \;.
\end{eqnarray}
This archetypical model has a number of important 
applications, e.g., the Kondo problem, 
interstitial tunneling in solids \cite{leggett}, 
quantum computing \cite{garg}, and electron transfer
reactions \cite{david}, to mention only a few. 
The bare two-level system (TLS) has a tunneling matrix element $\Delta$
and the asymmetry (bias) $\epsilon$ between the two localized energy levels  
($\sigma_x$ and $\sigma_z$ are 
Pauli matrices). Dissipation is introduced via a linear heat bath, i.e.,
an arbitrary collection of harmonic oscillators $\{x_\alpha\}$
bilinearly coupled to $\sigma_z$.
Concerning the TLS dynamics, all information about the coupling 
to the bath is contained in the spectral density
$J(\omega)=(\pi/2) \sum_\alpha (c^2_\alpha/m_\alpha\omega_\alpha)\,
\delta (\omega-\omega_\alpha)$,
which has a quasi-continuous form in typical condensed-phase applications.
$J(\omega)$ dictates the form of 
the (twice-integrated) bath correlation function 
($\beta=1/k_B T$),
\begin{equation}
Q(t) = \int_0^\infty \frac{d\omega}{\pi \hbar} 
\frac{J(\omega)}{\omega^2}
\, \frac{\cosh[\omega\hbar\beta/2]-
\cosh[\omega(\hbar\beta/2-it)]}{\sinh[\omega\hbar\beta/2]}\;.
\end{equation}
For the calculations here, we assume an ohmic spectral density of the form
$J(\omega) =2\pi\hbar\alpha \omega \exp(-\omega/\omega_c)$,
for which $Q(t)$ can be found in closed form \cite{egger94}.  Here
$\omega_c$ is a cutoff frequency, and the damping strength is 
measured by the dimensionless Kondo parameter $\alpha$. 
In the scaling limit $\Delta\ll \omega_c$,
and assuming $\alpha<1$,
all dependence on $\omega_c$ enters via a renormalized
 tunnel splitting \cite{leggett}
\begin{equation}
\Delta_{\rm eff} 
= [\cos(\pi \alpha)\Gamma(1-2\alpha)]^{1/2(1-\alpha)}
(\Delta/\omega_c)^{\alpha/(1-\alpha)}  \Delta \;,
\end{equation}
and powerful analytical \cite{leggett,saleur} and alternative numerical methods 
\cite{stockburger,makri} are
available for computing the nonequilibrium dynamics. 

At this point some remarks are in order.
Basically all other published numerical methods except
real-time PIMC can deal
only with equilibrium quantities, see, e.g., Refs.\cite{costi,voelker},
or explicitly introduce approximations  \cite{stockburger,makri,winterstetter,song}.
Regarding the latter class, 
mostly  Markovian-type approximations concerning the time-range of the 
interactions introduced by the influence functional have been implemented.
Our approach is computationally more expensive than other methods 
\cite{stockburger,makri,costi,voelker,winterstetter,song},
but at the same time it is unique in yielding numerically
exact results for the nonequilibrium spin-boson dynamics for arbitrary
bath spectral densities.
It is particularly valuable away from the scaling
regime where important applications, e.g., coherent (nonequilibrium) electron 
transfer reactions in the adiabatic regime,  
are found but basically all other methods fail
to yield exact results.  
Finally we briefly compare the present approach to our 
previously published PIMC method \cite{egger94}.  For not exceedingly
small $\alpha$, it turns out that the latter method is just equivalent
to the $K=1$  limit of the present method.  From Table I and 
the discussion below, it is thus apparent that MLB is significantly
more powerful in allowing for a study of much longer real times than
previously.  

We study the quantity $P(t) =  \langle \sigma_z(t) \rangle$
under the nonequilibrium initial preparation $\sigma_z(t<0)=+1$.
$P(t)$ gives the time-dependent difference of the 
quantum-mechanical occupation probabilities of the left and right states,
with the particle initially confined to the left state.
To obtain $P(t)$ numerically, we take the discretized
path-integral representation of Ref.\cite{egger94} and 
trace out the bath to get a long-ranged effective action, the 
``influence functional''.  In discretized form
the TLS path is represented by spins $\sigma_i,\sigma'_i=\pm 1$
on the forward- and backward-paths, respectively.
The total action $S$ consists of three terms.
First, there is the ``free'' action $S_0$ determined by
 the bare TLS propagator $U_0$, 
\begin{equation}
\exp(-S_0)= \prod_{i=0}^{P-1} U_0(\sigma_{i+1},\sigma_i;t^*/P)
\; U_0 (\sigma'_{i+1},\sigma'_i;-t^*/P) \;.
\end{equation}
The second is the influence functional, $S_I= S_I^{(1)} + S_I^{(2)}$, which 
contains the long-ranged interaction among the spins,
\begin{eqnarray}
S_I^{(1)} &=& \sum_{j\geq m} (\sigma_j-\sigma'_j) 
\Bigl\{ L'_{j-m} \, (\sigma_m-\sigma'_m) \\ \nonumber &+& i L^{''}_{j-m}\,
(\sigma_m+\sigma'_m) \Bigr\} \;,
\end{eqnarray}
where $L_j=L'_j+iL_j^{''}$ is given by \cite{egger94}
\begin{equation}
L_{j}  =  [Q((j+1)t^*/P)+Q((j-1) t^*/P) - 2Q(jt^*/P) ]/4
\end{equation}
for $j>0$, and $L_{0}=Q(t^*/P)/4$. 
In the scaling regime at $T=0$, this effective action has
interactions $\sim \alpha/t^2$ between the spins
(``inverse-square Ising model'').  
The contribution  
\begin{equation}
S_I^{(2)} = i(t^*/P) \sum_m  \gamma(m t^*/P) (\sigma_m-\sigma'_m) 
\end{equation}
gives the interaction with the imaginary-time branch [where $\sigma_z=+1$],
where the damping kernel 
\begin{equation}
\gamma(t)= \frac{2}{\pi \hbar} \int_0^\infty
d\omega \frac{J(\omega)}{\omega} \,\cos(\omega t) \;.
\end{equation}
For clarity, we focus on the most difficult case of an unbiased
two-state system at zero temperature, $\epsilon=T=0$.
 To ensure that the Trotter error
is negligibly small, we have systematically increased $P$
for fixed $t^*$ until convergence was reached. 
Typical CPU time requirements per $10^4$ MC samples are
4 hours for $P=26, L=2, K=1000$, or 6 hours for 
$P=40, L=3, K=600$, where 
the simulations were carried out on SGI Octane workstations.
The memory requirements for these two
cases are 60 Mbyte and 160 Mbyte, respectively.
Data were collected from several $10^5$ samples. 

For $\alpha=0$, the bare TLS dynamics $P(t)=\cos(\Delta t)$ is
accurately reproduced.  As mentioned before, the performance 
is slightly inferior to the original MLB approach \cite{mlbrt} 
which is now applicable due
to the absence of the influence functional and the associated
long-ranged interactions.  Turning to the situation where
a bath is present, we first study  the case
 $\alpha=1/2$ and $\omega_c/\Delta=6$. 
 The exact $\alpha=1/2$ result \cite{leggett},
$P(t) = \exp(-\Delta_{\rm eff} t)$, valid in the scaling regime
$\omega_c/\Delta\gg 1$, was accurately reproduced, indicating that 
the scaling regime is reached already for moderately large
$\omega_c/\Delta$.  Typical parameters used in the MLB simulations
and the respective average sign are listed in Table~\ref{table1}.
The first line in Table~\ref{table1} corresponds to the naive 
approach.  For $\alpha=1/2$, it turns out that our previous 
PIMC scheme \cite{egger94} yields a comparable performance
to the $K=1$ version of this MLB method. It is then
clear from Table~\ref{table1} that the average sign and
hence the signal-to-noise ratio can
be dramatically improved thus allowing for a study of
significantly longer timescales $t^*$ than before.
For a fixed number of levels $L$, the average sign 
grows by increasing the parameter $K$.  Alternatively, for  
fixed $K$, the average sign increases with $L$.  Evidently,
the latter procedure is more efficient in curing the sign problem,
but at the same time computationally more expensive.  In practice,
it is then necessary to find a suitable compromise.

\begin{figure}
\epsfxsize=1.2\columnwidth
\epsffile{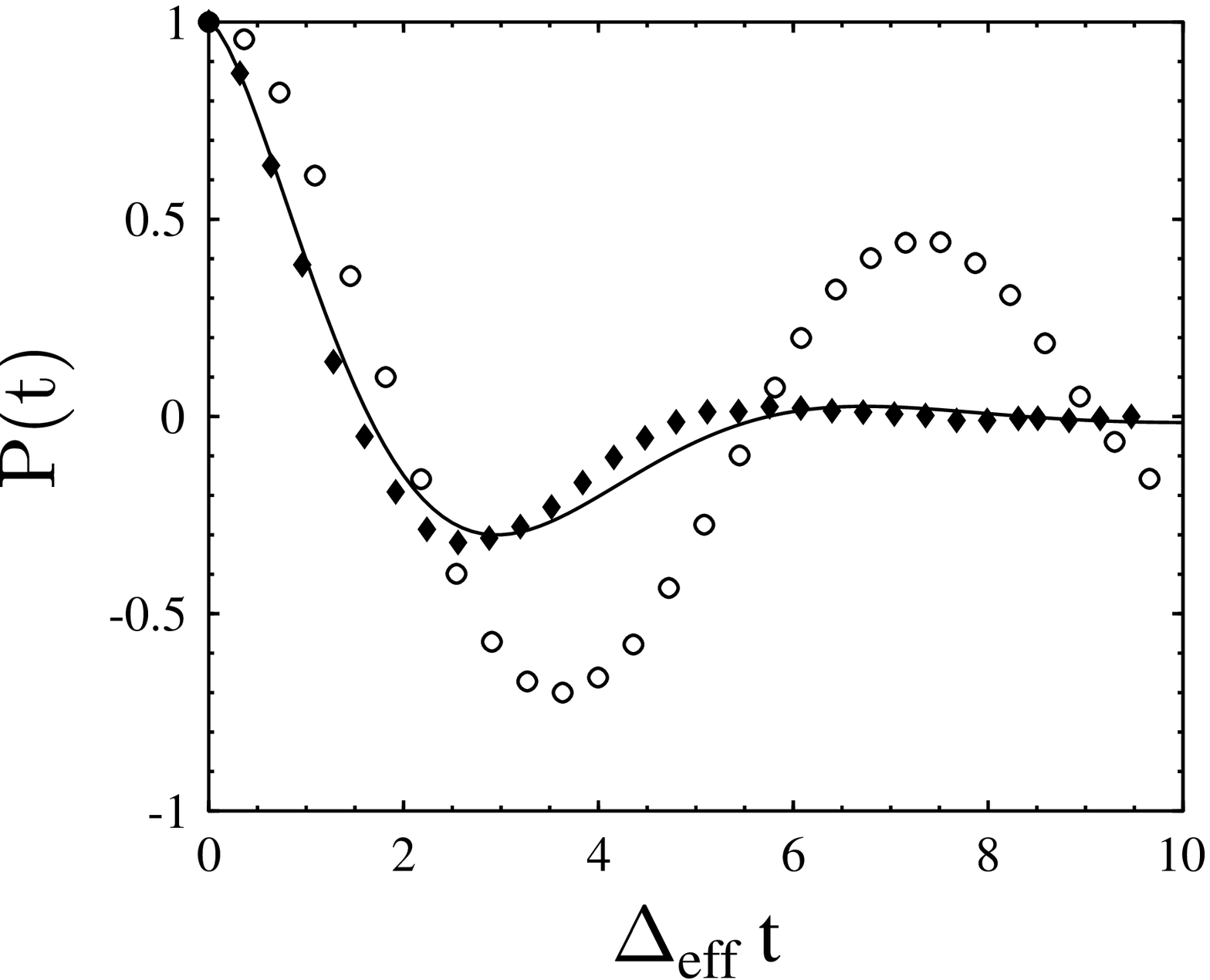}
\caption[]{\label{figure1} Scaling curves for
$P(t)$ for $\alpha=1/4$ with
$\omega_c/\Delta=6$ (closed diamonds) and  
$\omega_c/\Delta=1$ (open circles).  The solid curve is the
NIBA prediction. The approach of Ref.\cite{egger94}
 becomes unstable for $\Delta_{\rm eff} t>4$ in both cases. 
Statistical errors are of the order of the symbol sizes.}
\end{figure}

Figure \ref{figure1} shows scaling curves for $P(t)$ at $\alpha=1/4$ for 
$\omega_c/\Delta=6$ and $\omega_c/\Delta=1$. According to the
$\alpha=1/2$ results, $\omega_c/\Delta=6$
is expected to be within the scaling regime.
This is confirmed by a comparison to
the noninteracting blip approximation (NIBA)
\cite{leggett}.  The minor deviations of the NIBA curve from
the exact result are in accordance with Refs.\cite{egger94,saleur}
for $\alpha\leq 1/2$.
However, for $\omega_c/\Delta=1$, scaling concepts
(and also NIBA) are expected to fail even qualitatively.  
Clearly, the MLB results show that away from the scaling region,
quantum coherence is able to persist for much longer, and both frequency and 
decay rate of the oscillations differ significantly from the predictions
of NIBA.  In electron transfer reactions in the  
adiabatic-to-nonadiabatic crossover regime, 
such coherence effects can then strongly influence the low-temperature
dynamics.  One obvious and important consequence of these coherence effects is the
breakdown of a rate description, implying that theories based on 
an imaginary-time formalism might not be appropriate in this regime.
A detailed analysis of this crossover regime
using MLB is currently in progress.
 
\section{Conclusions}
\label{conc}

In this paper, we have extended the multilevel blocking (MLB)
approach of Refs.\cite{mlb,mlbrt}
to path-integral Monte Carlo simulations with long-ranged effective actions 
along the Trotter direction.  
For clarity, we have focussed on real-time simulations here,
but believe  that a similar approach can also be helpful in many-fermion
computations,  e.g., in auxiliary-field fermion simulations of lattice 
fermions.   The practical usefulness of the approach was demonstrated
by computing the nonequilibrium real-time dynamics of the dissipative
two-state system.  Here the effective action (influence functional) 
arises by integrating out the linear heat bath.  For a heat bath of the
ohmic type, at $T=0$ the corresponding interactions among different
time slices decay only with a slow inverse-square power law.
 
In the present implementation of MLB, the basic blocking idea 
operates on multiple time scales by carrying out a subsequent sampling 
at longer and longer times.  During this procedure,
the interference information collected at shorter times is
 taken fully into account without invoking any approximation.
Under such an approach, at the expense of large memory requirements,
the severity of the sign
problem can be significantly relieved.
The proposed approach allows to 
study time scales not accessible to previous real-time 
path-integral simulations for the spin-boson system.

\acknowledgements

We wish to thank M. Dikovsky and J. Stockburger for useful 
discussions.
This research has been supported by the Volkswagen-Stiftung,
by the National Science Foundation under Grants No.~CHE-9257094
and No.~CHE-9528121, by the Sloan Foundation, and by the 
Dreyfus Foundation.

\begin{table}
\caption[]{\label{table1}
MLB performance for $\alpha=1/2$, $\omega_c/\Delta=6$, 
$\Delta t^*=10$, $P=40$, and several $L$.
$q_\ell$ denotes the number of slices for $\ell=1,\ldots L$. }
\begin{tabular}{llll}
$K$   & $L$ & $q_\ell$        & $\langle {\rm sgn} \rangle$ \\ \hline
1     & 1   & 40              & 0.03  \\   
200   & 2   & 30 - 10         & 0.14  \\
800   & 2   & 30 - 10         & 0.20  \\
200   & 3   & 22 - 12 - 6     & 0.39  \\
600   & 3   & 22 - 12 - 6     & 0.45  \\
\end{tabular}
\end{table}


\begin{references}
\bibitem{foot1}
The calculation of thermal correlation functions is possible
after minor modifications.
\end{references}
\end{document}